\documentclass[aps,prl,twocolumn,floats,balancelastpage,showpacs,showkeys,preprintnumbers,floatfix,nofootinbib,superscriptaddress]{revtex4}
\usepackage{graphicx,amsmath,amssymb,amsfonts, soul,color}
\usepackage{hyperref}

\begin{document}
\title{Search for $\gamma$-ray line signals from dark matter annihilations\\ in the inner Galactic halo from ten years of observations with H.E.S.S.}

\author{H.E.S.S. Collaboration}
\noaffiliation

\author{H.~Abdallah} 
\affiliation{Centre for Space Research, North-West University, Potchefstroom 2520, South Africa}

\author{A.~Abramowski}
\affiliation{Universit\"at Hamburg, Institut f\"ur Experimentalphysik, Luruper Chaussee 149, D 22761 Hamburg, Germany}

\author{F.~Aharonian}
\affiliation{Max-Planck-Institut f\"ur Kernphysik, P.O. Box 103980, D 69029 Heidelberg, Germany}
\affiliation{Dublin Institute for Advanced Studies, 31 Fitzwilliam Place, Dublin 2, Ireland}
\affiliation{National Academy of Sciences of the Republic of Armenia, Marshall Baghramian Avenue, 24, 0019 Yerevan, Republic of Armenia}

\author{F.~Ait Benkhali}
\affiliation{Max-Planck-Institut f\"ur Kernphysik, P.O. Box 103980, D 69029 Heidelberg, Germany}

\author{E.O.~Ang{\"u}ner}
\affiliation{Instytut Fizyki J\c{a}drowej PAN, ul. Radzikowskiego 152, 31-342 Krak{\'o}w, Poland}

\author{M.~Arakawa}
\affiliation{Department of Physics, Rikkyo University, 3-34-1 Nishi-Ikebukuro, Toshima-ku, Tokyo 171-8501, Japan}

\author{M.~Arrieta}
\affiliation{LUTH, Observatoire de Paris, PSL Research University, CNRS, Universit\'e Paris Diderot, 5 Place Jules Janssen, 92190 Meudon, France}

\author{P.~Aubert}
\affiliation{Laboratoire d'Annecy-le-Vieux de Physique des Particules, Universit\'{e} Savoie Mont-Blanc, CNRS/IN2P3, F-74941 Annecy-le-Vieux, France}

\author{M.~Backes}
\affiliation{University of Namibia, Department of Physics, Private Bag 13301, Windhoek, Namibia}

% ...

\begin{abstract}
Spectral lines are among the most powerful signatures for dark matter (DM) annihilation searches in very-high-energy $\gamma$ rays. The central region of the Milky Way halo is one of the most promising targets 
given its large amount of DM and proximity to Earth.  We report on a search for a monoenergetic spectral line from self-annihilations of DM particles in the energy range from 300~GeV to 70~TeV using a two-dimensional maximum likelihood method taking advantage of both the spectral and spatial features of signal versus background.
The analysis makes use of Galactic Center (GC) observations accumulated over ten years (2004 - 2014) 
 with the H.E.S.S. array of ground-based Cherenkov telescopes. No significant $\gamma$-ray excess above the background is found. We derive upper limits on the annihilation cross section $\langle \sigma v\rangle$ for monoenergetic DM lines at the level of $\rm 4\times10^{-28} cm^3s^{-1}$ at 1~TeV, assuming an Einasto DM profile for the Milky Way halo. For a DM mass of 1~TeV, they improve over the previous ones by a factor of six. 
The present constraints are the strongest obtained so far for DM particles in the mass range 300 GeV - 70~TeV. 
Ground-based $\gamma$-ray observations have reached sufficient sensitivity to explore 
relevant velocity-averaged cross sections for DM annihilation into two $\gamma$-ray photons at the level expected from the thermal relic density for TeV DM particles.
 \end{abstract}


\maketitle 

\section{Introduction}
\label{sec:introduction}  
Cosmological measurements show that about 85\% of the matter in the universe is non-baryonic cold dark matter (DM)~\cite{Adam:2015rua}. A leading class of DM particle candidates consists of weakly interacting massive particles (WIMPs)~\cite{Jungman:1995df,Bergstrom:2000pn,Bertone:2004pz,Feng:2010gw}. Thermally produced in the early universe, stable particles with mass and coupling strength at the electroweak
scale have a relic density which is consistent with that of observed DM. In dense DM regions, the self-annihilation of WIMPs would give rise today to Standard Model particles, including a possible emission of very-high-energy (VHE, E$_{\gamma}$ $\gtrsim$ 100 GeV) $\gamma$ rays in the final state. 

DM self-annihilations are expected to produce a continuum spectrum of $\gamma$ rays up to the DM mass $m_{\rm DM}$ from prompt annihilation into quarks, heavy leptons or gauge bosons\footnote{A secondary emission from inverse Compton scattering and bremsstrahlung of electrons produced in the decay chain.}, and $\gamma$-ray lines. 
While the continuum signal is non-trivial to distinguish from other standard broadband astrophysical emissions, the DM self-annihilation at rest into $\gamma X$ with $X = \gamma, h, Z$ or a non Standard Model neutral particle, would give a prominent and narrow spectral line at an energy $E_{\gamma} = m_{\rm DM}(1-m_X^2/4m_{\rm DM}^2)$, only limited by the detector resolution given the low ($\sim$10$^{-3}$c) relative velocity of the DM particles. When DM self-annihilates into charged particles, additional $\gamma$ rays are present from final state radiation and virtual internal bremsstrahlung. This produces bumpy bremsstrahlung features giving a wider line that peaks at an energy near $m_{\rm DM}$~\cite{Bergstrom:1989jr,Bringmann:2007nk}.

Since the DM is strongly constrained to be electrically neutral, the annihilation into monoenergetic $\gamma$ rays is typically loop-suppressed compared to the continuum signal, and the velocity-weighted annihilation cross section into two photons is about $10^{-2} - 10^{-4}$ of the total velocity-weighted annihilation cross section $\langle\sigma v\rangle$ (see, for instance, Refs.~\cite{Bergstrom:1997fh,Ferrer:2006hy,Gustafsson:2007pc,Profumo:2008yg}). For WIMPs produced in a standard thermal history of the Universe, $\langle\sigma v\rangle$ is about $3 \times 10^{-26}$ cm$^{-3}$s$^{-1}$ in order to reproduce the observed density of DM in the universe~\cite{Steigman:2012nb}. VHE $\gamma$-ray lines can be detected by ground-based Cherenkov telescope arrays such as
H.E.S.S. (High Energy Stereoscopic System). 

The central region of the Galactic halo observed in VHE $\gamma$ rays is among the most compelling targets to search for monoenergetic line signals from DM annihilations due to its proximity to Earth and predicted large DM concentration. For WIMPs in the TeV mass range, the strongest constraints so far reach $\langle \sigma v\rangle\sim$ 3$\times$10$^{-27} $cm$^3$s$^{-1}$ at 1~TeV~\cite{Abramowski:2013ax} using four years of observations of the Galactic Center (GC) region with H.E.S.S. 

The energy differential $\gamma$-ray flux produced by the annihilation of self-conjugate DM particles of mass $m_{\rm DM}$ in a solid angle $\rm \Delta\Omega$ can be written as
\begin{equation}
\begin{aligned}
\label{eq:promptflux}
\frac{{\rm d} \Phi}{{\rm d} E_{\gamma}} (E_{\gamma}, \Delta\Omega) = \frac{\langle \sigma v \rangle}{8\pi \ m_{\rm DM}^2} 
 \frac{{\rm d} N}{{\rm d} E_{\gamma}}(E_{\gamma})\times J(\Delta\Omega) \ ,  \\
{\rm with} \, \, \, \,  J(\Delta\Omega)=\int_{\Delta\Omega}\int_{\rm LOS}{\rm d} s \ {\rm d} \Omega \ \rho^2(r(s,\theta))\ . 
 \end{aligned} 
\end{equation} 
The first term includes the DM particle physics properties. 
${\rm d} N/{\rm d} E_{\gamma}(E_{\gamma}) =  2 \delta(m_{\rm DM} - E_{\gamma})$ is the differential $\gamma$-ray yield per annihilation into two photons. $J(\Delta\Omega)$ denotes the integral of the square of the DM density $\rho$ along the line of sight (LOS) in a solid angle $\Delta\Omega$. It is commonly referred to as the {\it J-factor}~\cite{Bergstrom:1997fj}. The coordinate $r$ is defined by $r = (r^2_{\odot}+s^2-2 r_{\odot} s\ {\rm cos}\ \theta)^{1/2}$, where $s$ is the distance along the line of sight, and $\theta$ is the angle between the direction of observation and the GC.  $r_{\odot}$ is the distance of the observer with respect to the GC, taken equal to 8.5~kpc~\cite{Ghez:2008ms}. In this work, we consider DM density distributions parametrized by cuspy profiles, for which archetypes are the Einasto~\cite{Springel:2008by} and Navarro-Frenk-White (NFW)~\cite{Navarro:1996gj} profiles (See, also, Ref.~\cite{Cirelli:2010xx}). Cored profiles are not studied here since they need specific data taking and analysis procedures to be probed as shown in Ref.~\cite{HESS:2015cda}.

From ten years of observations of the GC region with the initial four telescopes of H.E.S.S., 
we present here a new search for DM annihilations into monoenergetic narrow $\gamma$-ray lines\footnote{We consider as a monoenergetic narrow line each structure that is narrow on the scale of the 10\% energy resolution of H.E.S.S.} in the inner Galactic halo~\cite{Abramowski:2013ax}. 
Exploiting the increased photon statistics, we perform the search in the mass range 300~GeV - 70~TeV with an improved technique for $\gamma$-ray selection and reconstruction and a two-dimensional (2D) likelihood-based analysis method using the spectral and spatial features of the DM annihilation signal with respect to background. 

\section{Data analysis}
\label{sec:analysis}
The dataset was obtained from GC observations with H.E.S.S. phase I  during the years 2004 - 2014 as in
Ref.~\cite{Abdallah:2016ygi} with telescope pointing positions  between  0.5$^{\circ}$  to 1.5$^{\circ}$ from the GC. 
Standard criteria for data quality selection are applied to the data to select $\gamma$-ray events~\cite{Aharonian:2006pe}. In addition, observational zenith angles higher than 50$^{\circ}$ are excluded to minimize systematic
uncertainties in the event reconstruction. The dataset amounts to 254 h (live time) with a mean zenith angle of the selected observations of 19$^{\circ}$. The $\gamma$-ray event selection and reconstruction make use of an advanced  
semi-analytical shower model technique~\cite{2009APh32231D} in order to determine the direction and the energy of each event. With this technique, the energy resolution defined as the distribution of $\Delta E/E = (E_{\rm reco}-E_{\rm true})/E_{\rm true}$ has a r.m.s of 10\% above 300 GeV.
 This technique is also very well suited to mitigate the effects expected from the variations of the night sky background (NSB) in the field of view~\cite{2009APh32231D}. In the GC region, broad NSB variations may induce systematic effects in the event acceptance and, therefore in the normalization of the signal and background region exposure~\cite{2013APh4117D,HESS:2015cda}. A discussion on the systematic effects from NSB variations in the present analysis is given in Ref.~\cite{supplement}.

The search for a DM signal is performed in regions of interest (ROIs) defined as annuli with inner radii of 0.3$^{\circ}$ to 0.9$^{\circ}$ in radial distance from the GC, and width of 0.1$^{\circ}$, hereafter referred to as the ON region. Following Ref.~\cite{Abdallah:2016ygi}, a band of $\pm$0.3$^{\circ}$ along the Galactic plane is excluded to 
avoid astrophysical background contamination from the VHE sources such as HESS J1745-290 coincident in position with the supermassive black hole Sagittarius A*~\cite{Aharonian:2004wa,Aharonian:2009zk}, the supernova/pulsar wind nebula G0.9+0.1~\cite{Aharonian:2005br}, and a diffuse emission extending along the Galactic plane~\cite{Aharonian:2006au,HESS2014sla,Abramowski:2016mir}. A disk with 0.4$^{\circ}$ radius masks the supernova remnant HESS J1745-303~\cite{Aharonian:2008gw}.

\begin{figure}[!ht]
\centering
\includegraphics[width=0.45\textwidth]{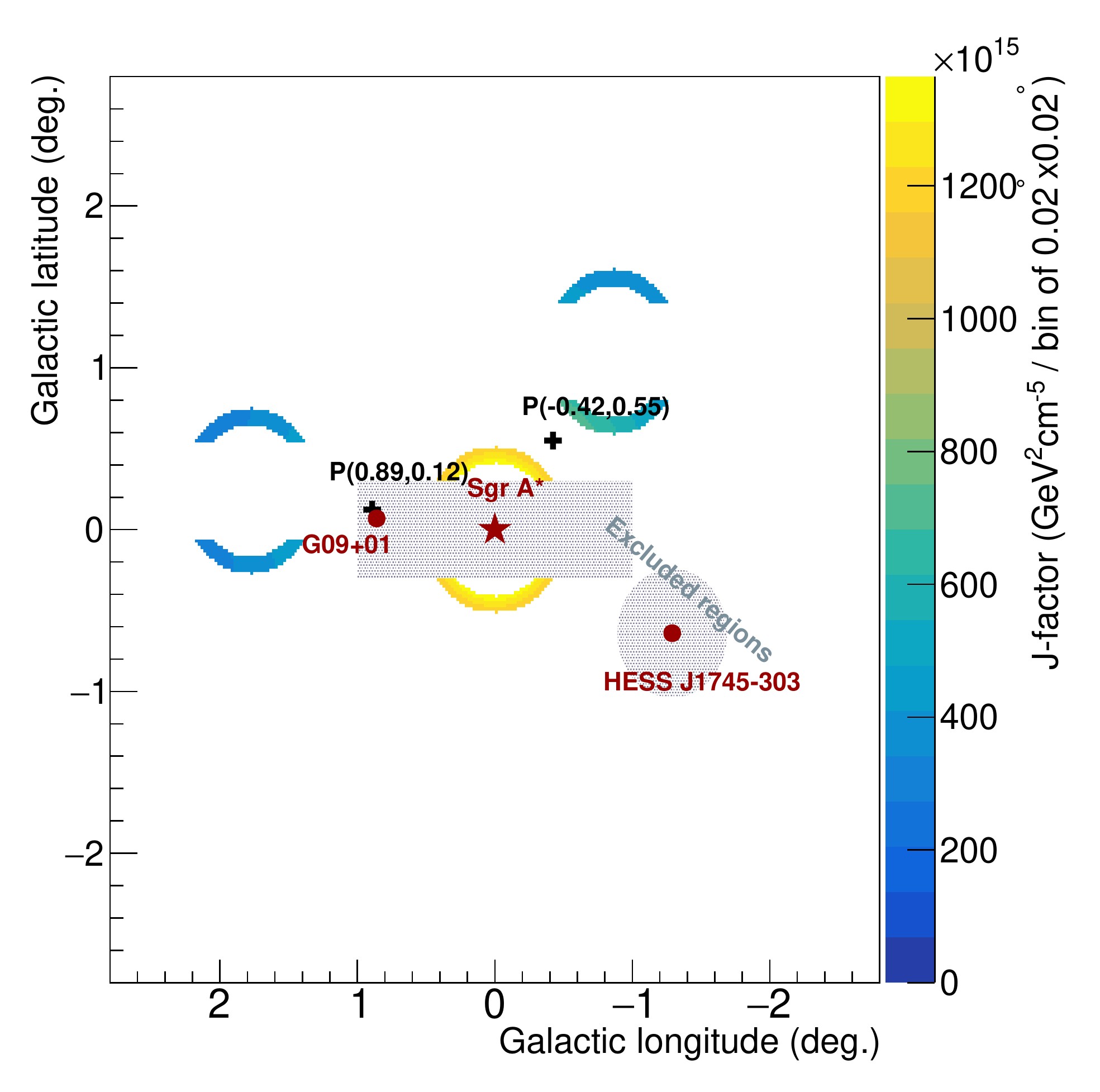}
\caption{Schematic of the background measurement technique for the ROI 2 and two different telescope pointing positions, in Galactic coordinates. The OFF region is taken symmetrically to the ON region from a given observational pointing position (black cross). Two OFF regions are shown, each one corresponding to a specific pointing position.
ON and OFF regions have the same angular size and shape. The positions of Sgr A* (red star), G0.9+0.1 (red dot) and HESS J1745-303 (red dot) are shown. The grey-filled box with Galactic latitudes from $-$0.3$^{\circ}$ to +0.3$^{\circ}$ and the grey-filled disc are excluded for signal and background measurements. The color scale gives the J-factor value per spatial bin of 0.02$^{\circ} \times$0.02$^{\circ}$ for the Einasto DM profile.}
\label{fig:Fig0}
\end{figure}

The background events are selected for each observation in an OFF region chosen symmetrically to the ON region with respect to the observational pointing position. The ON and OFF regions are thus taken with same acceptance and observation conditions, and have the same shape and solid angle size as shown in Fig.~1 in the Supplemental Material~\cite{supplement}. Such a measurement technique enables an accurate background determination which does not require further acceptance correction. The OFF regions are always sufficiently far away from the ON region to obtain a significant DM gradient between the ON and OFF regions for cuspy DM profiles. For such profiles, we consider OFF regions which are expected to contain always fewer DM events than the ON regions. 
Fig.~\ref{fig:Fig0} shows an example of J-factor values in the ON and OFF regions for the ROI 2 and two specific telescope pointing positions. For the pointing position P(0.89,0.12), a gradient of about 3.5 is obtained between the ON and OFF regions. See Supplemental Material~\cite{supplement} for more details, which 
includes Ref.~\cite{Abramowski:2011hc}.

We perform a 2D binned Poisson maximum likelihood analysis in order to exploit the spatial and spectral characteristics of the DM signal with respect to background. The energy range is divided into 60 logarithmically spaced bins between 300 GeV and 70 TeV. Seven spatial bins corresponding to ROIs defined as the above-mentioned annuli of 0.1$^{\circ}$ width are chosen following Ref.~\cite{Abdallah:2016ygi}. For a given DM mass, the total likelihood function is obtained from the product of the individual Poisson likelihoods $\mathcal{L}_{\rm ij}$ over the spatial bins $i$ and the energy bins $j$,
\begin{widetext}
\begin{equation}
\mathcal{L}_{\rm ij}(\mathbf{N_{\rm ON}},\mathbf{N_{\rm OFF},\mathbf{\alpha}}|\mathbf{N_{\rm S}},\mathbf{N_{\rm S}'},\mathbf{N_{\rm B}})=\frac{(N_{\rm S,ij}+N_{\rm B,ij})^{N_{\rm ON,ij}}}{N_{\rm ON,ij}!}e^{-(N_{\rm S,ij}+N_{\rm B,ij})}\frac{(N_{\rm S,ij}'+\alpha_i N_{\rm B,ij})^{N_{\rm OFF,ij}}}{N_{\rm OFF,ij}!}e^{-(N_{\rm S,ij}'+\alpha_i N_{\rm B,ij})}.
\label{eq:likelihood}
\end{equation}
\end{widetext}
For each bin $(i,j)$, $N_{\rm ON}$ and $N_{\rm OFF}$ are the measured number of events in the ON and OFF regions, respectively. $\alpha=\Delta\Omega_{\rm OFF}/\Delta\Omega_{\rm ON}$ corresponds to the ratio of the solid angle sizes of the OFF and ON regions. Here, $\alpha_i$ = 1 by definition of the ON and OFF regions. 
The expected number of background events $N_{\rm B}$ in the ON region is extracted from residual background measurements in the dataset.
$N_{\rm S}$ and $N_{\rm S}'$ stand for the number of signal events expected in the ON and OFF regions, respectively.  They are obtained by folding the theoretical number of DM events with the energy-dependent acceptance and energy resolution of H.E.S.S. for this data set. The $\gamma$-ray line signal is 
represented by a Gaussian function at the line energy $E_{\gamma} = m_{\rm DM}$ with a width of $\sigma/E_{\gamma}$. 
The vectors ${\bf N_{\rm ON}}$, ${\bf N_{\rm OFF}}$, ${\bf N_{\rm S}}$, ${\bf N_{\rm S}'}$, ${\bf N_{\rm B}}$, and ${\boldsymbol \alpha}$ represent the lists of the corresponding quantities for all bins.

In the absence of statistically significant $\gamma$-ray excess in the ON region with respect to the OFF region, constraints on the DM line flux and velocity-weighted annihilation cross section can be obtained from the likelihood ratio test statistic given by $ {\rm TS}=-2 \ln(\mathcal{L}(m_{\rm DM},\langle { \sigma v} \rangle)/\mathcal{L}_{\rm max}(m_{\rm DM},\langle { \sigma v} \rangle))$. In the high statistics limit, TS follows a $\chi^2$ distribution with one degree of freedom~\citep{Rolke:2004mj}. Values of $\Phi$ and $\langle \sigma v \rangle$ for which the TS value is higher than 2.71 provide one-sided 95\% confidence level (C.L.) upper limits on the flux and velocity-weighted annihilation cross section, respectively.
Uncertainties in the energy reconstruction scale and the energy resolution affect these limits by less than 25\%.
 The systematic uncertainty arising from NSB variations in the field of view modify the limits up to 60\%. See Ref.~\cite{supplement} for more details.

\section{Results}
\label{sec:results}
We find no statistically significant $\gamma$-ray excess in any of the ROIs with respect to the background.
A cross-check analysis using independent event calibration and reconstruction~\cite{Parsons:2014voa} confirms the absence of any significant excess.
\begin{figure*}[ht!]
\centering
\includegraphics[width=0.45\textwidth]{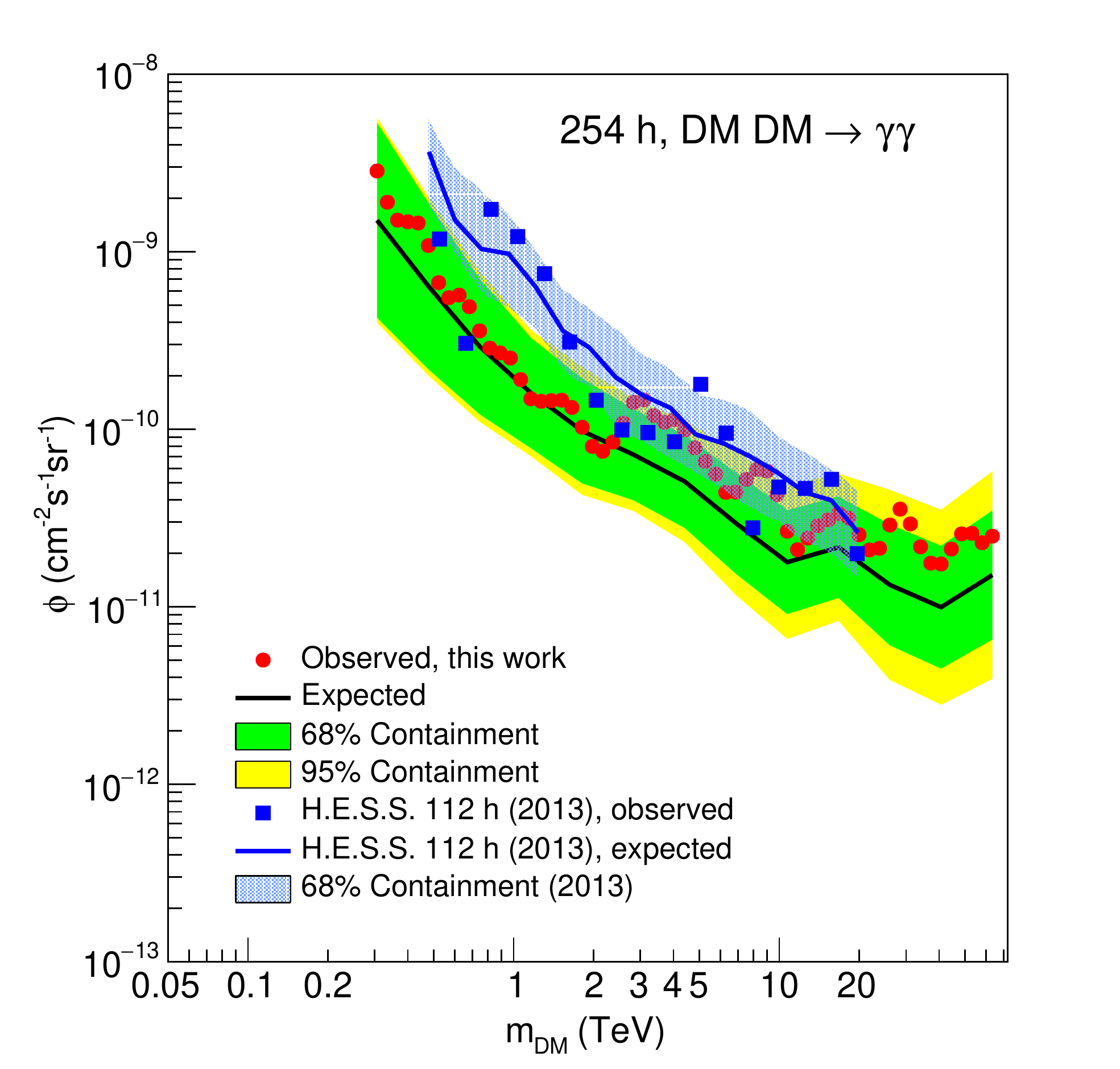}
\includegraphics[width=0.45\textwidth]{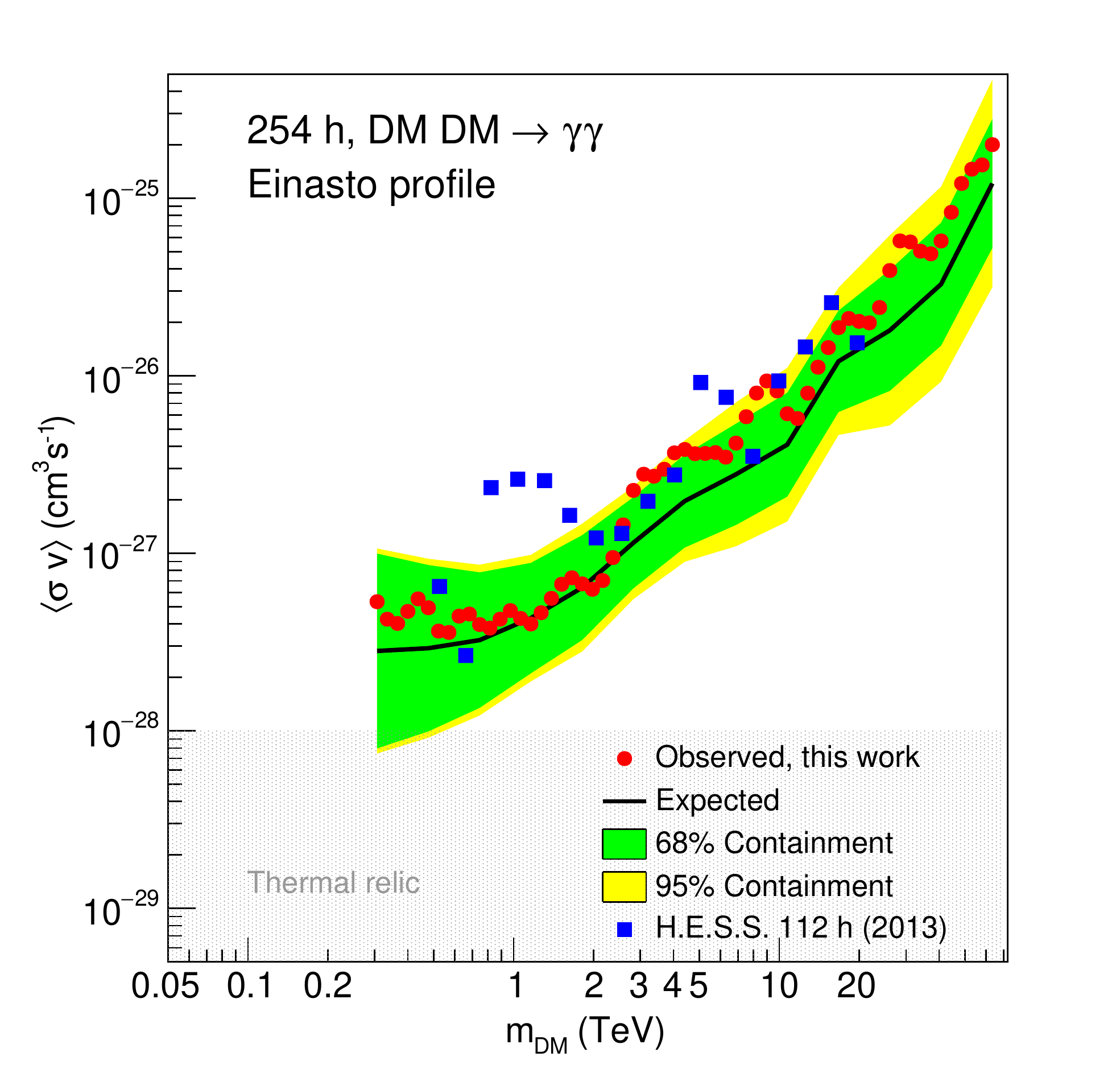}
\caption{Constraints on the flux $\Phi$ (left panel) and on the velocity-weighted annihilation cross section $\langle \sigma v \rangle$ (right panel) for the prompt annihilation into two photons derived from H.E.S.S. observations taken over ten years (254~h of live time) of the inner 300 pc of the GC region. The constraints are expressed in terms of 95\% C.~L. upper limits as a function of the DM mass m$_{\rm DM}$ for the Einasto profile.
The observed limits are shown as red dots. Expected limits are computed from 1000 Poisson realizations of the expected background derived from blank-field observations at high Galactic latitudes. The mean expected limit (black solid line) together with the 68\% (green band) and 95\% (yellow band) C.~L. containment bands are shown. The bands include the statistical and the systematic uncertainties.
The observed limits derived in the analysis of four years (112 h of live time) of GC observations by H.E.S.S.~\cite{Abramowski:2013ax} are shown as blue squares, together with the mean expected limit (blue solid line) and the 68\% containment band (blue shaded area) in the left panel.
The natural scale for monochromatic $\gamma$-ray line signal is highlighted as a grey-shaded area in the right panel.}
\label{fig:results}
\end{figure*}
\begin{figure}[ht!]
\includegraphics[width=0.45\textwidth]{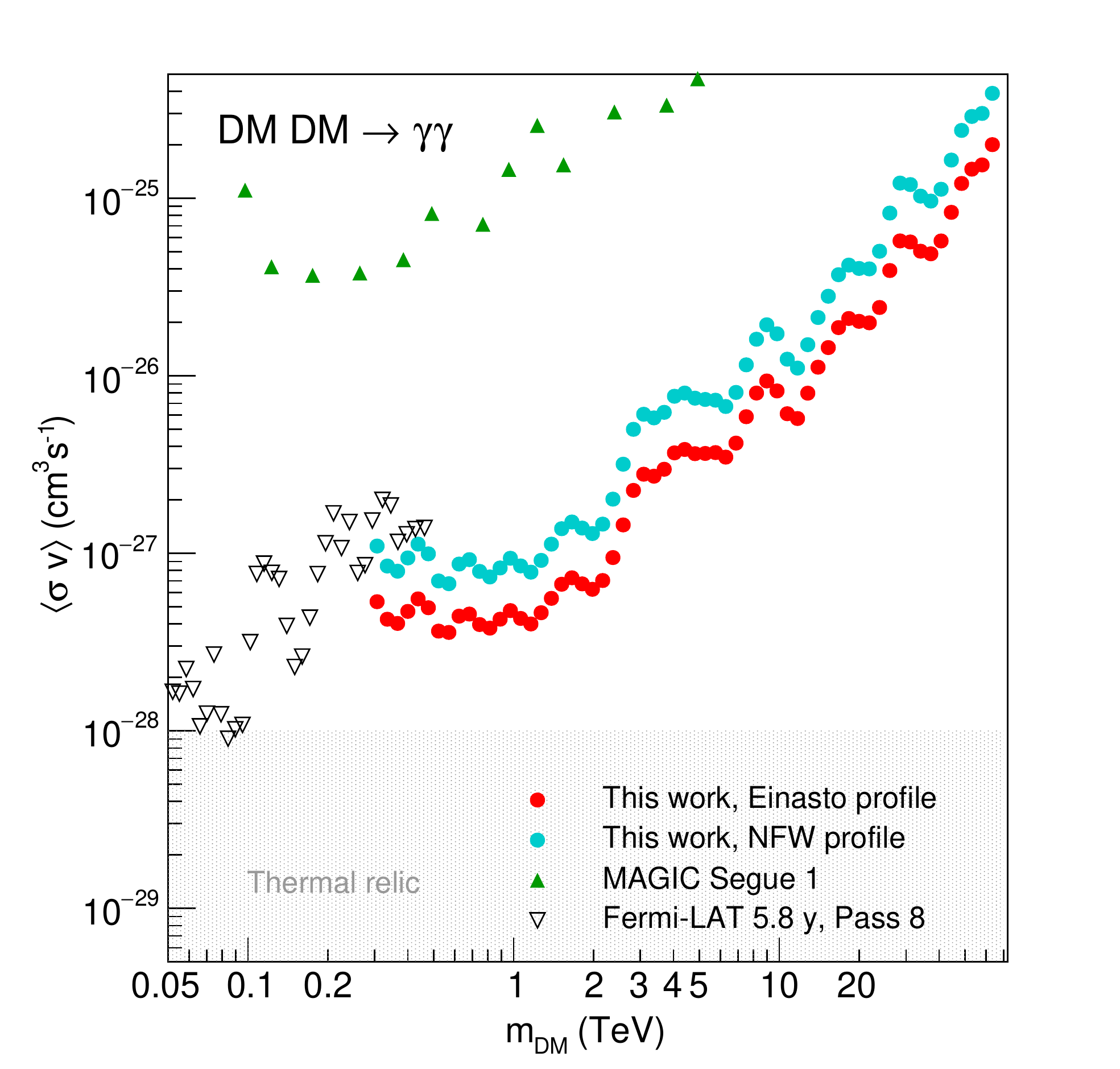}
\caption{Comparison of constraints for prompt annihilation into two photons obtained by H.E.S.S. for the Einasto (red dots) and NFW (cyan dots) profiles, respectively, with the limits from the observations of the Milky Way halo by Fermi-LAT~\cite{Ackermann:2015lka} (black triangles) as well as the limits from 157 hours of MAGIC observations of the dwarf galaxy Segue~1~\cite{Aleksic:2013xea} (green triangles). The grey-shaded area shows the natural scale for a monochromatic $\gamma$-ray line signal. 
}
\label{fig:summary}
\end{figure} 
We derive upper limits on $\Phi$ and $\langle \sigma v \rangle$  at 95\% C.~L. for DM masses from 300~GeV to 70~TeV. The left panel of Fig.~\ref{fig:results} shows the observed upper limits at 95\% C.~L. on the flux from prompt DM self annihilations into two photons for the Einasto profile\footnote{Assuming a kpc-sized cored DM density distribution  such as the Burkert profile would weaken the limits by about two-to-three orders of magnitude.}. In order to check that the observed limits are in agreement with random fluctuations of the expected background, we computed expected limits using the likelihood ratio TS from 1000 Poisson realizations of the expected background derived from observations of blank fields at high latitudes where no signal is expected (see Supplemental Material~\cite{supplement}). 
For each DM mass, the mean expected upper limit and the 68\% and 95\% containment bands are extracted from the obtained $\Phi$ and $\langle \sigma v \rangle$ distributions, and are plotted on the left panel of Fig.~\ref{fig:results}. In addition to the statistical uncertainty, the containment bands include the systematic uncertainties coming from the energy scale, the energy resolution and NSB variations in the field of view~\cite{supplement}.

We obtain the largest improvement in the observed flux limits compared to the previous results published in Ref.~\cite{Abramowski:2013ax} for a DM particle mass of 1~TeV, where the limits are stronger by a factor of 6.
The improved photon statistics, 
the likelihood analysis method using both ON and OFF Poisson terms and the 2D likelihood analysis method yield
an increase of sensitivity by a factor of about 1.4, 1.8 and 1.3, respectively.
The remaining improvement factor comes from the improved $\gamma$-ray event selection and reconstruction technique used in the present analysis~\cite{2009APh32231D}. The 95\% C.~L. observed flux limit reaches  $\sim$1.6$\times$10$^{-10}$ cm$^{-2}$s$^{-1}$sr$^{-1}$ at 1~TeV. The right panel of Fig.~\ref{fig:results} shows the 95\% C.~L. upper limits on $\langle \sigma v \rangle$ for the Einasto profile, together the natural scale for gamma-ray  line signals\footnote{The upper bound is expected for $\gamma$-ray lines from thermal Higgsinos annihilating into two photons~\cite{Bergstrom:1997fh}, and from final state radiation and internal bremmstrahlung~\cite{Bringmann:2007nk}.}. The observed and the expected limits together with their 68\% and 95\% C.~L. containment bands are plotted. 
For a DM particle of mass 1~TeV, the observed limit is~$\langle \sigma v \rangle \simeq$~4$\times$10$^{-28}$cm$^3$s$^{-1}$.

In Fig.~\ref{fig:summary}, we show a comparison of our results with the current constraints on the prompt DM self-annihilation into two photons obtained from 5.8 years of observations of the Milky Way halo\footnote{The observed limits for Fermi-LAT are extracted for the DM density profile labelled as Einasto R16 in Ref.~\cite{Ackermann:2015lka}.} by the Fermi-LAT satellite~\cite{Ackermann:2015lka} and the limits from 157 hours of observations of the dwarf galaxy Segue 1\footnote{The J-factor of Segue 1 used in Ref.~\cite{Aleksic:2013xea} could be overestimated by a factor of 100 as shown in Ref.~\cite{Bonnivard:2015xpq}.} by the MAGIC ground-based Cherenkov telescope instrument~\cite{Aleksic:2013xea}. The previous limits obtained by H.E.S.S. from 112 hours of observations of the GC~\cite{Abramowski:2013ax} are also plotted.

\section{Summary and discussion}
We presented a new search for monoenergetic VHE $\gamma$-ray lines from ten years of observation of the GC (254 h of live time) by the phase I of H.E.S.S. with a novel statistical analysis technique using a 2D maximum likelihood method. No significant $\gamma$-ray excess is found and we exclude a velocity-weighted annihilation cross section into two photons of $\rm 4\times 10^{-28}$ cm$^3$s$^{-1}$ for DM particles with a mass of 1 TeV for an Einasto profile. We obtain the strongest limits so far for DM masses above 300 GeV. 

The limits obtained in this work significantly improve over the strongest constraints so far from 112 hours of H.E.S.S. observations towards the GC region in the TeV mass range~\cite{Abramowski:2013ax}. The new constraints cover a DM mass range from 300 GeV up to 70 TeV. They provide a significant mass range overlap with the Fermi-LAT constraints. They surpass the Fermi-LAT limits by a factor of about four for a DM mass of 300 GeV~\cite{Ackermann:2015lka}.

Despite the gain in sensitivity, our upper limits are still larger than the typical cross sections  for thermal WIMPs at $\langle \sigma v \rangle \sim 10^{-29}$cm$^3$s$^{-1}$ expected for supersymmetric neutralinos~\cite{Bergstrom:1997fh}. However, there are several WIMP models which predict larger cross sections. While being not thermally produced, they still produce the right relic DM density. Among the wide class of heavy WIMP models, those with enhanced $\gamma$-ray lines (see, for instance, Ref.~\cite{Hisano:2003ec}) are in general strongly constrained by the results presented here. The present results can be applied to models with wider lines  while dedicated analyses taking into account the intrinsic line shapes are required. They include models with $\gamma$-ray boxes~\cite{Ibarra:2015tya}, scalar~\cite{Giacchino:2015hvk} and Dirac~\cite{Duerr:2015wfa} DM models, as well as the canonical Majorana DM triplet fermion known as the {\it Wino} in Supersymmetry~\cite{Baumgart:2017nsr}. 

The limits obtained by H.E.S.S. in this work are complementary to the ones obtained from direct detection and collider production ({\it i.e.}, LHC) searches. While the latter ones are powerful techniques to look for DM of masses of up to about hundred GeV, the indirect detection with $\gamma$-rays carried out with Fermi-LAT satellite and ground-based Cherenkov telescopes is the most powerful approach to probe DM in the higher mass regime, as shown from several studies developed in the framework of effective field theory~\cite{Charles:2016pgz} and, more recently, using the simplified-model approaches (see, for instance, Ref.~\cite{Sirunyan:2017hci}). Observations with ground-based Cherenkov telescopes such as H.E.S.S. are unique to probe multi-TeV DM through the detection of $\gamma$-ray lines.

The upcoming searches with H.E.S.S. towards the inner Galactic halo will exploit additional observations including the fifth telescope at the center of the array. Since 2014, a survey of the inner galaxy is carried out with the H.E.S.S. instrument focusing in the inner 5$^{\circ}$ of the GC. This survey will allow us to probe a larger source region of DM annihilations and alleviate the impact of the uncertainty of the DM distribution in the inner kpc of the Milky Way on the sensitivity to DM annihilations. A limited dataset ($\sim$15 hours) of this survey using 2014 observations with the fifth telescope only was used to constrain the presence of a 130 GeV DM line in the vicinity of the GC~\cite{Abdalla:2016olq}. 
Observations including the fifth telescope will allow us to probe DM lines down to 100 GeV. In addition, 
a higher fraction of stereo events in the energy range from hundred to several hundred GeV is expected
from the increased number of stereo triggers between the fifth telescope and one of the recently-upgraded smaller telescopes. Beyond the sensitivity improvement expected from increased photon statistics, the inner galaxy survey will provide a larger fraction of photons in regions of devoid of known standard astrophysical emissions, therefore 
of prime interest for DM searches. Within the next few years DM searches with H.E.S.S. will enable an even more in-depth exploration of the WIMP paradigm for DM particles in the hundred GeV to ten TeV mass range.

\section{Acknowledgments}
The support of the Namibian authorities and of the University of Namibia in facilitating the construction and operation of H.E.S.S. is gratefully acknowledged, as is the support by the German Ministry for Education and Research (BMBF), the Max Planck Society, the German Research Foundation (DFG), the Alexander von Humboldt Foundation, the Deutsche Forschungsgemeinschaft, the French Ministry for Research, the CNRS-IN2P3 and the Astroparticle Interdisciplinary Programme of the CNRS, the U.K. Science and Technology Facilities Council (STFC), the IPNP of the Charles University, the Czech Science Foundation, the Polish National Science Centre, the South African Department of Science and Technology and National Research Foundation, the University of Namibia, the National Commission on Research, Science \& Technology of Namibia (NCRST), the Innsbruck University, the Austrian Science Fund (FWF), and the Austrian Federal Ministry for Science, Research and Economy, the University of Adelaide and the Australian Research Council, the Japan Society for the Promotion of Science and by the University of Amsterdam. We appreciate the excellent work of the technical support staff in Berlin, Durham, Hamburg, Heidelberg, Palaiseau, Paris, Saclay, and in Namibia in the construction and operation of the equipment. This work benefited from services provided by the H.E.S.S. Virtual Organisation, supported by the national resource providers of the EGI Federation.

\bibliography{bibl}

\clearpage
\appendix
\setcounter{equation}{0}
\widetext
\begin{center}
{\bf \large \large Supplemental Material: Search for $\gamma$-ray line signals from dark matter\\ annihilations towards the in the Galactic halo from ten years of observations with H.E.S.S.}
\end{center}



\section{Observational dataset at the Galactic Center and definition of the regions of interest}
Optimal observation strategies target regions where a high density of DM and reduced astrophysical $\gamma$-ray background are expected. The Galactic Centre (GC) region is among the best target to look for line emission from DM annihilations. 
The  GC dataset is obtained from about 600 observational runs with pointing positions taken up to 1.5$^{\circ}$ in radial distance from the GC with the four telescopes of the phase I of H.E.S.S. The dataset is taken with all the telescopes pointed in the same direction in the sky and requires at least that two telescopes are triggered by an event. Given the field of view of H.E.S.S.-I observations and the pointing positions for GC observations, useful photon statistics is obtained up to about 3.5$^{\circ}$.

Given the field of view of H.E.S.S.-I observations and the pointing positions, we search for the dark matter (DM) signal in a circular region of $1^\circ$ centered at the GC. 
In order to take into account the different spatial behavior of signal and background,  the region of interest is further divided in 7 regions of interest (ROI) defined as rings of width 0.1$^\circ$, with inner radii from 0.3$^\circ$ to 0.9$^\circ$. They are hereafter referred to as ON regions. Interestingly, the signal-to-noise  ratio (S/N) assuming a cuspy DM distribution as described by the Einasto and NFW profiles, maximizes at about 1$^\circ$ from the GC. However, the S/N distribution as a function of the angular distance from GC has a smooth dependency from one degree up to a few degrees.

\section{Signal and background measurement}
\begin{figure*}[!ht]
\centering
\includegraphics[width=0.45\textwidth]{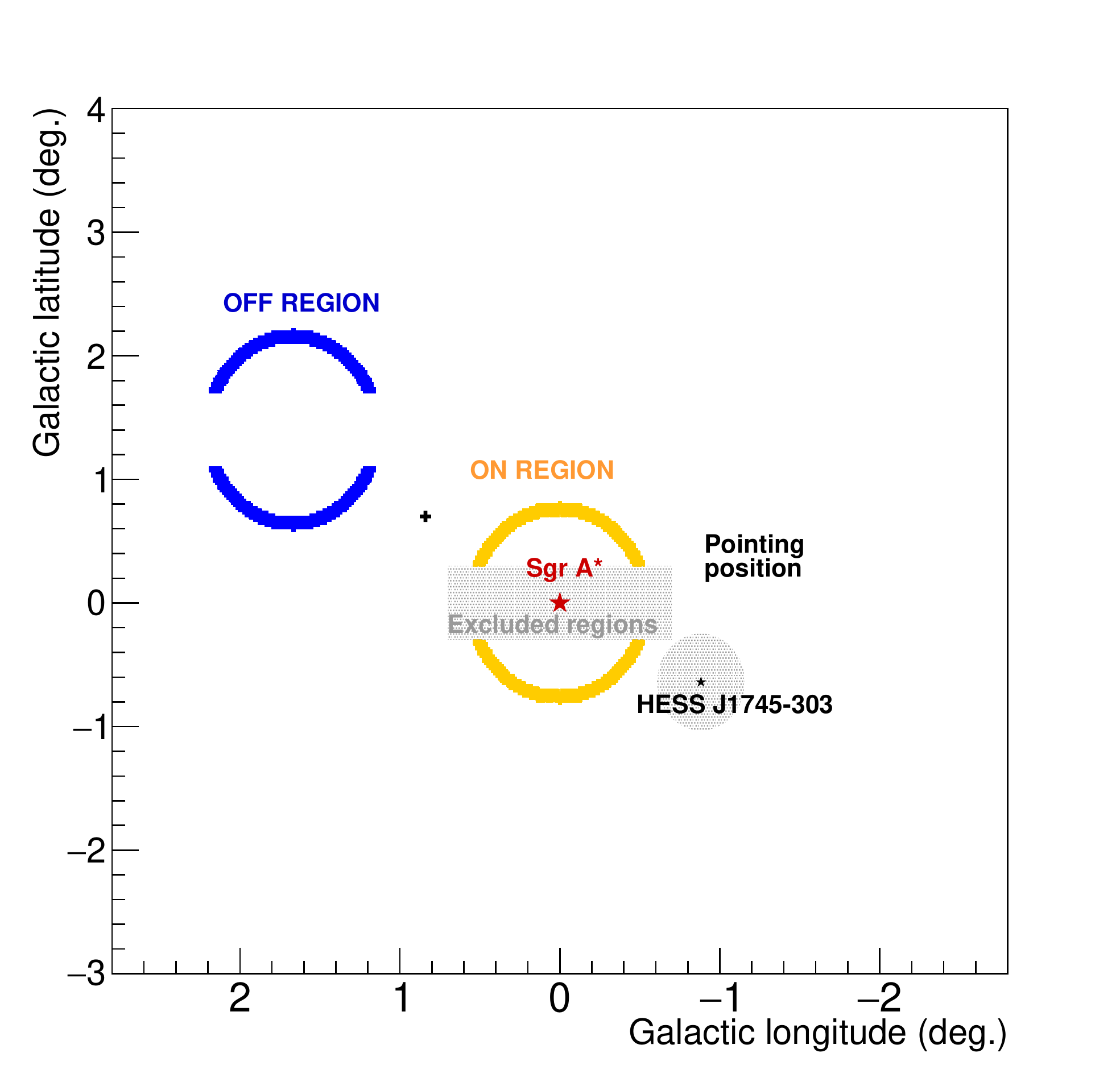}
\caption{Geometrical construction of the background region (OFF region) in Galactic coordinates for a specific pointing position and one ROI (ON region). The OFF region (blue open ring) is determined as the symmetrical region of the ON region (yellow open ring) from the pointing position (black cross). The ON region considered here is the ROI 5. The red and back stars mark the position of Sagittarius A$^\ast$ and the source HESS J1745-303, respectively. The grey-shaded box-like region corresponds to the excluded region along the Galactic plane. Part of the ring that intersects the box is excluded from the ON region, and therefore from the OFF region by symmetry. The grey-filled disc corresponds to the excluded region for HESS J1745-303.}
\label{fig:geometrical}
\end{figure*}

The background for a given ROI is measured for each pointing position of the GC observations with H.E.S.S. I, in the same field of view as for the signal, in an OFF region taken geometrically symmetric to the ON region with respect to the pointing position of the observation as in Ref.~\cite{Abramowski:2011hc}. For each pointing direction $\boldsymbol n_{\rm p}$, the ON and OFF regions are constructed for each ROI. For all directions $\boldsymbol n$ in a given ROI, the direction $\boldsymbol n'$ = 2($\rm \boldsymbol n \cdot \boldsymbol n_{\rm p}) \boldsymbol n_{\rm p} - \boldsymbol n$ is computed, which is reflected with respect to the pointing direction. The directions $\boldsymbol n$ and $\boldsymbol n'$ are added to the ON region and the OFF region, respectively, if none of them falls inside an excluded region and if $\boldsymbol n$ points closer to the GC than $\boldsymbol n'$. The thus obtained ON and OFF regions have the same shapes and solid angles. Such a technique provides an accurate background measurement in an OFF region taken under the same observational and instrumental conditions as for the signal measurement in the ON region. With this background measurement technique, the ON and OFF-source data are taken in the same high voltage or mirror reflectivity conditions because ON and OFF data are taken simultaneously. The main regions of the sky with VHE $\gamma$-ray sources are marked as excluded regions, in order to avoid pollution of the DM search by astrophysical processes. Fig.~\ref{fig:geometrical} shows the geometrical construction of an OFF region (blue) for a given ON region (yellow) chosen here to be ROI 5. The pointing position is marked by a black cross.  

\begin{figure*}[!ht]
\centering
\includegraphics[width=0.45\textwidth]{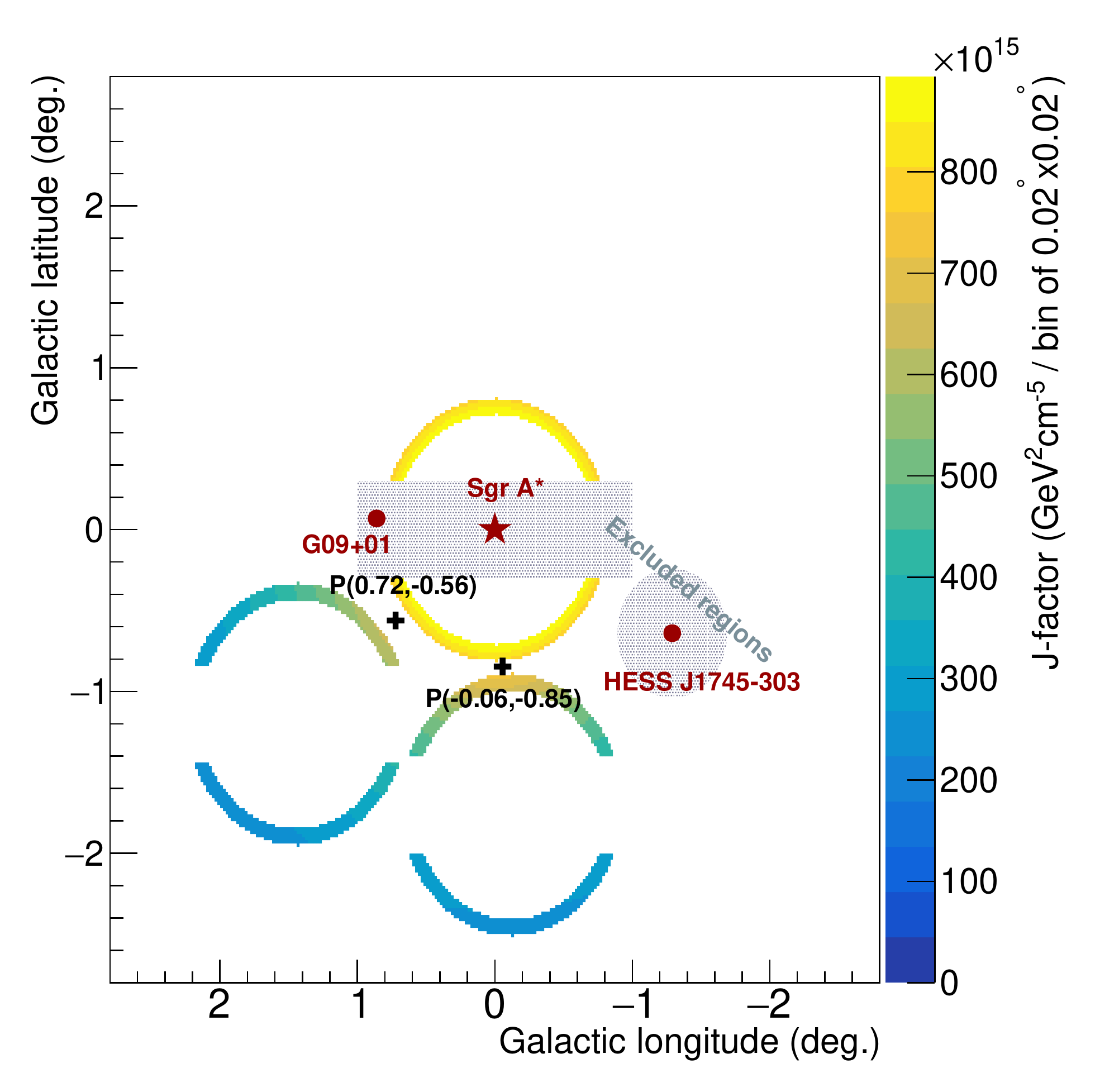}
\caption{Schematic of the background measurement technique for the ROI 5 with two pointing positions, in Galactic coordinates. The OFF region is taken symmetrically to the ON region from a given observational pointing position (black cross). Two OFF regions are shown, each one corresponding to a specific pointing position.
By construction, ON and OFF regions have the same angular size and shape. The positions of Sgr A* (red star), G0.9+0.1 and HESS J1745-303 (red dots) are shown. The grey-filled box with Galactic latitudes from $-$0.3$^{\circ}$ to +0.3$^{\circ}$ and the grey-filled disc are excluded for signal and background measurements. The color scale gives the J-factor value per spatial bin of 0.02$^{\circ} \times$0.02$^{\circ}$ size.}
\label{fig:background_method}
\end{figure*}

The expected DM signal in the ON and OFF regions is computed using the J-factor values integrated in each region. Examples of the J-factor values in 0.02$^{\circ} \times$0.02$^{\circ}$ spatial bins are shown in Fig.~\ref{fig:background_method} for ON and OFF regions using two pointing positions, in Galactic coordinates.
The OFF regions corresponding to the ON region ROI 5 are plotted for the pointing positions P(0.72,$-$0.56) and P($-$0.06,$-$0.85), respectively. For the former pointing position, the J-factor ratio between the ON and OFF regions is $\sim$2.3, while for the latter it is $\sim$2.5. In principle, OFF regions at small galactic latitudes ($|b|<$3$^{\circ}$) are affected by the diffuse $\gamma$-ray emission along the Galactic plane~\cite{Abramowski:2014vox}. However, the contribution to the measured OFF fluxes was estimated to be at most at the few percent level and therefore negligible.

\section{Expected background determination and expected limit computation}
The expected limits are computed from blank-field observations at high Galactic latitude ($|b| > 10^{\circ}$). 
In these extragalactic observations, all VHE sources are excluded 
to construct a $\gamma$-ray background map
 free from VHE $\gamma$-ray source emissions. From this database, the residual background is derived in the 
 same observation conditions as for the GC dataset: for each run, we select from the database the background corresponding to the zenith angle, mirror efficiency and offset of the run. This procedure is repeated for all the runs. It provides the mean expected background map in the inner 1$^{\circ}$ for each energy bin, and consequently in all considered ROIs. It is hereafter referred to as the expected OFF distributions in the ROIs. In addition, the expected OFF distributions are weighted to account for the excluded regions used in the observed background determination. 1000 Poisson realizations of the expected OFF distribution and of the expected signal are created for each of the seven ROIs. For each realization, the likelihood ratio test statistic is used for each mass to derive the $\langle \sigma v \rangle$ value corresponding TS =  2.71. From the computed $\langle \sigma v \rangle$ distribution, the mean of the distribution together the 68\% and 95\% containment bands are extracted. 

\section{Systematic uncertainties on the expected limit}
In the GC region, the Night Sky Background (NSB) level is subject to strong variations due to the presence of bright stars in the field of view. It varies from about 150 MHz up to 300 MHz on a spatial scale of one degree. 
A careful treatment of the NSB in the GC region is carried out using the shower template method as described in Ref.~\cite{2009APh32231D} where the contribution of the NSB noise in every pixel of the camera is
modelled. This method does not require any image cleaning to extract the pixels illuminated by the shower. In addition, the stability of the semi-analytical shower model analysis against missing pixels in the event reconstruction (dead pixels or switch-off pixels because of bright stars) has been proven and compared to the more standard Hillas analysis. 

From regions of the sky with relatively low NSB of about 100 MHz measured in northern-east quadrant to regions with NSB as high as 300 MHz as seen in the southern-west quadrant, a systematic effect of 4\% (1$\sigma$) in
the $\gamma$-ray rate is found. Adding 4\% more events in the OFF PDFs shifts the mean expected limit as a function of the DM mass from a few percents up to 60\%.

Given the above-mentioned background measurement technique on a run-by-run basis, the ratio between the ON and OFF exposures is one. However, a systematic uncertainty may arise from the assumption of azimuthal symmetry in the field of view. For a given telescope pointing position, we stacked all the gamma-ray events falling in a ring of inner and outer radii of 0.5$^{\circ}$ and 0.6$^{\circ}$, respectively. 
The azimuthal distributions of events in this ring are compatible within the expected statistical fluctuations and no systematic effect is found. 

In order to estimate the impact of the systematic uncertainty in the energy scale on the limit, we artificially increase and decrease the value of the reconstructed energy of each gamma-ray event by 10\%. This shifts the mean expected limit up to 15\%. The energy resolution for the H.E.S.S.-I instrument is of 10\% above gamma-ray energies of 300 GeV  in the analysis method used here~\cite{2009APh32231D}. An artificial deterioration of the energy resolution from 10 to 20\% implies a shift in the expected limit by 25\%.

\section{Dark matter halo profile parametrizations}
The DM distribution is assumed to follow a cuspy distribution for which the NFW and Einasto profiles are archetypal parametrizations. The profile parametrisations are given for the Einasto and NFW profiles, respectively, by:
\begin{equation}
\label{eq:profiles}
\rho_{\rm E}(r) = \rho_{\rm s}  \exp \left[-\frac{2}{\alpha_{\rm s}}\left(\Big(\frac{r}{r_{\rm s}}\Big)^{\alpha_{\rm s} }-1\right)\right]\\
\quad {\rm and} \quad \rho_{\rm NFW}(r) = \rho_{\rm s}\left(\frac{r}{r_{\rm s}}\Big(1+\frac{r}{r_{\rm s}}\Big)^2\right)^{-1}  \ , 
\end{equation}
assuming a local DM density of $\rho_{\odot} = 0.39\ \rm GeV\ cm^{-3}$. 
Table~\ref{tab:tab1} gives the parameters of the profiles used here.
\begin{table}[h]
\centering
\begin{tabular}{c|c|c|c}
Profiles & Einasto & NFW & Einasto~\cite{Cirelli:2010xx}  \\
\hline
\hline
$\rho_{\rm s}$ (GeVcm$^{-3}$) & 0.079 & 0.307 &  0.033 \\
$r_{\rm s}$ (kpc) 		        & 20.0   & 21.0   & 28.4  \\
$\alpha_{\rm s}$                        & 0.17   &   /       &  0.17  \\
\hline
\hline
\end{tabular}
\caption{Parameters of the cuspy profiles used for the DM distribution. The Einasto and NFW profiles considered here follow Ref.~\cite{Abramowski:2011hc}. An alternative normalization of the Einasto profile~\cite{Cirelli:2010xx} is also used. \label{tab:tab1}}
\end{table}
Table~\ref{tab:tab2} gives the inner and outer radii  for each ROI together with its solid angle size and its J-factor for the Einasto profile. 
\begin{table}[h]
\centering
\begin{tabular}{c|c|c|c|c}
i$^{\rm th}$ ROI & Inner radius & Outer radius & $\Delta\Omega$& J-factor \\
 & [deg.] & [deg.] & [10$^{-5}$ sr] &  [10$^{20}$ GeV$^2$cm$^{-5}$] \\
\hline
\hline
1 & 0.3 & 0.4 & 3.1 & 4.1 \\
2 & 0.4 & 0.5 & 5.0 & 5.4 \\
3 & 0.5 & 0.6 & 6.9 & 6.3\\
4 & 0.6 & 0.7 & 8.8 & 7.0\\
5 & 0.7 & 0.8 & 10.7 & 7.5\\
6 & 0.8 & 0.9 & 12.6 & 8.0\\
7 & 0.9 & 1.0 & 14.5 & 8.3\\
\hline
\hline
\end{tabular}
\caption{J-factor values in units of 10$^{20}$ GeV$^2$cm$^{-5}$ for the 7 RoIs. The first four columns give the region number, the inner radius, the outer radius, and the size in solid angle for each RoI, respectively. The last column provides the J-factor values for the Einasto profile considered in this work. \label{tab:tab2}}
\end{table}
 
\begin{figure*}[!ht]
\centering
\includegraphics[width=0.45\textwidth]{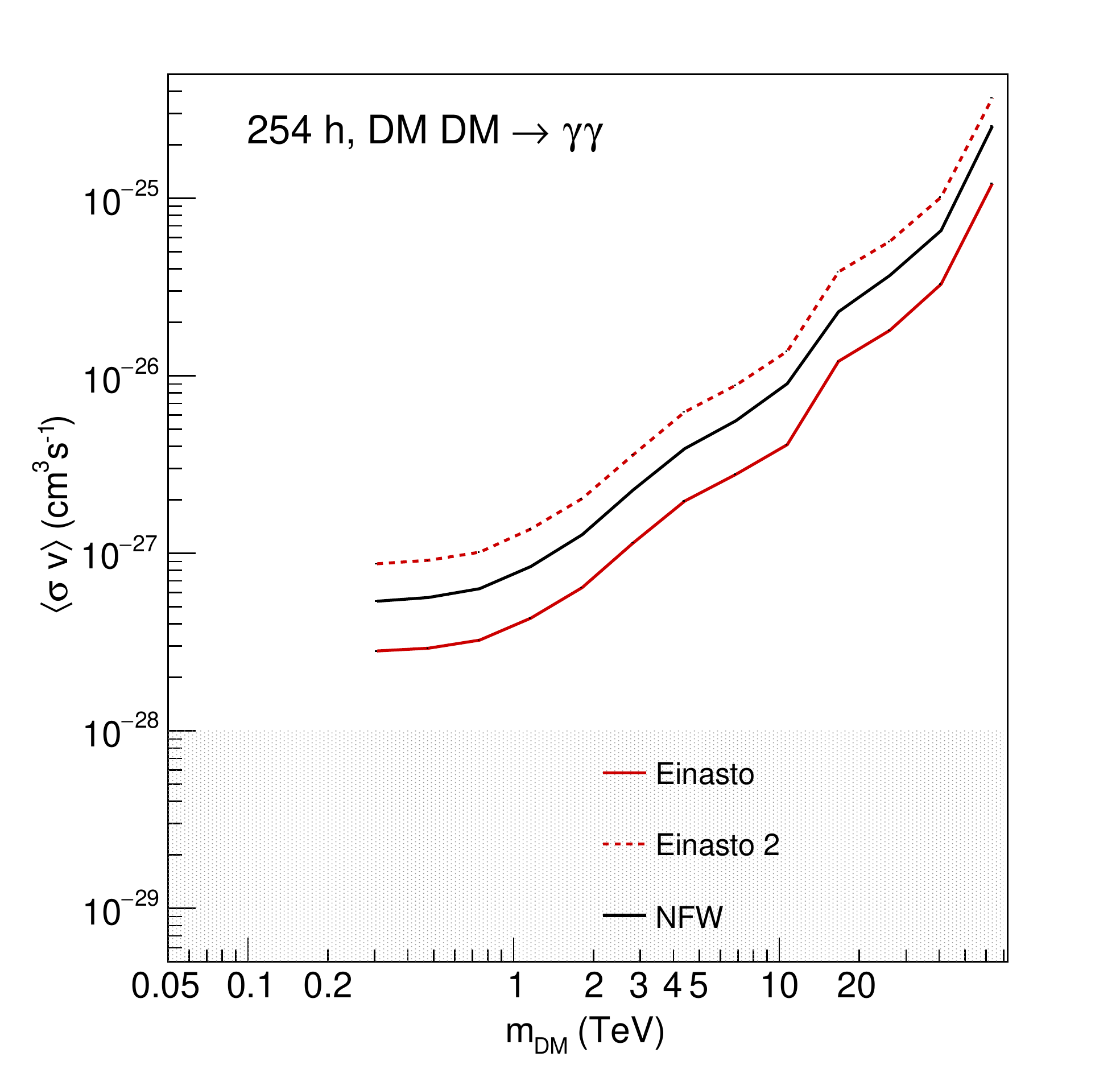}
\caption{Impact of the DM density distribution on the constraints on the velocity-weighted annihilation cross section $\langle \sigma v \rangle$ for the prompt annihilation into two photons. 
The mean expected limits at 95\% C. L. are shown as a function of the DM mass m$_{\rm DM}$ for the Einasto profile (solid red line), an alternative parametrization of the Einasto profile (dotted red line), and the NFW profile (solid black line), respectively.  
}
\label{fig:Fig7}
\end{figure*}

Figure~\ref{fig:Fig7} shows the impact of the DM distribution hypothesis on the 95\% C.L. mean expected limit for the NFW profile~\cite{Abramowski:2011hc} and an alternative parametrization of the Einasto profile extracted from Ref.~\cite{Cirelli:2010xx}. Their parameters of the DM profiles are given in Tab.~\ref{tab:tab1}. 

\end{document}